# Glass Transition of the Phase Change Material AIST and its Impact on Crystallization

*Julian Pries, Julia Sehringer, Shuai Wei, Pierre Lucas, Matthias Wuttig\**

J. Pries, J. Sehringer, Prof. M. Wuttig
    Institute of Physics IA, RWTH Aachen University, 52074 Aachen, Germany
    E-mail: wuttig@physik.rwth-aachen.de
Prof. S. Wei
    Department of Chemistry, Aarhus University, DK-8000 Aarhus-C, Denmark
Prof. P. Lucas
    Department of Materials Science and Engineering, University of Arizona, Tucson, AZ, 85721, United States

Keywords: Crystallization kinetics, glass transition temperature, phase change materials, metavalent bonding, structural relaxation, ultrafast differential scanning calorimetry (FDSC)

Engineering phase change materials (PCM) to realize superior data storage devices requires a detailed understanding of crystallization kinetics and its temperature dependence. The temperature dependence of crystallization differs distinctly between crystallizing from the glassy phase and the undercooled liquid (UCL). Hence, knowing the phase from which crystallization occurs is necessary for predicting the switching ability. Here, we measure the glassy dynamics and crystallization kinetics using calorimetry for heating rates spanning over six orders of magnitude. Our results show that the prominent PCM (Ag,In)-doped $Sb_2Te$ (AIST) exhibits a change from crystallizing from the glassy phase to crystallizing from the UCL at a critical heating rate of 5,000 K/s. Above the glass transition, the activation energy of crystallization changes drastically enabling rapid crystallization at elevated temperatures.

# 1. Introduction

Phase change materials (PCMs) are characterized by their ability to rapidly switch between the two solid states, namely between the amorphous and crystalline phase [1-4]. The electrical and optical property contrast between the two phases can then be exploited for various data storage applications [2, 5]. The property contrast was recently explained by a change in chemical bond mechanism, i.e. covalent bonding in the amorphous phase [6, 7] and metavalent bonding in the crystalline state [8-10]. One key aspect in computer memory applications is the data transfer rate determined by the switching speed. While amorphization of PCMs is demonstrated to occur on the picosecond time scale [11], (re-)crystallization usually takes about several nanoseconds even in the fastest switching PCMs [11-15]. Thus the latter constitutes the rate limiting process for memory operations [16]. In order to increase the recrystallization speed, detailed knowledge of the crystallization process is necessary. Crystallization occurs by nucleation of crystallites and their subsequent growth. Both the nucleation rate $I$ and the crystal growth velocity $v_g$ are limited by the kinetic factor that is approximately inversely proportional to the viscosity $\eta$ [17-20], which is a function of temperature $T$. While the viscosity of a (undercooled) liquid usually displays super-Arrhenius (fragile) behavior, it shows Arrhenius behavior in the glassy phase [21-23]. This results in different temperature dependencies of the nucleation and growth kinetics – and therefore the crystallization peak temperature $T_p$ – in the glassy phase and the undercooled liquid (UCL). Moreover, unlike the UCL, the glassy phase undergoes aging caused by structural relaxation when annealed at temperatures below the glass transition temperature $T_g$ [21, 24]. This aging leads to an increase in viscosity [25], which therefore leads to a decreased crystallization speed from the glass while crystallization from the UCL should largely be unaffected. Hence, knowing the phase, from which the material crystallizes, is crucial for predicting crystallization behavior at higher temperatures or heating rates, where crystallization speed can be optimized, as well as determining whether there are any transient effects on the crystallization ability upon aging/storing.

To identify the amorphous phase the material crystallizes from, and to obtain the temperature dependence of crystallization, the thermal response of amorphous as-deposited and pre-annealed $Ag_4In_3Sb_{67}Te_{26}$ (AIST) films prepared by sputter deposition is characterized using the conventional differential scanning calorimetry (DSC) and ultrafast Flash DSC (FDSC) with heating rates spanning more than six orders of magnitude from $3.3\times10^{-2}$ K/s to $4.0\times10^{4}$ K/s. The (F)DSC data yields valuable insights on glass dynamics, crystallization kinetics and their different temperature dependencies. The results allow us to reveal the distinction in the crystallization behaviors from the amorphous phase. Pre-annealing produces changes in the crystallization behavior and the nucleation tendency as inferred from fluctuation electron microscopy (FEM). This technique supports the conclusion from (F)DSC that AIST crystallizes from the glassy phases at low heating rates. Finally, these results are also found to be consistent with an estimate of the standard glass transition temperature $T_g$ of AIST derived from an analysis of the enthalpy recovery exotherm.

## 2. Results

The dominant feature in (F)DSC measurements of AIST is the crystallization exotherm that peaks at the temperature $T_p$. The heating rate dependence of $T_p$ is commonly represented in a Kissinger plot as shown in **Figure 1**, according to the Kissinger equation [26-29]

$$\ln\left(\frac{\vartheta}{T_p^2}\cdot \text{Ks}\right) = -\frac{E_k}{k_B T_p} + \ln\left(K_0 \frac{k_B}{E_K}\cdot \text{sK}\right) \tag{1}$$

where $E_k$ is the Kissinger activation energy of crystallization, $k_B$ is the Boltzmann constant and $K_0$ the prefactor of the rate constant $K(T)$. Figure 1 shows that, for both as-deposited and pre-annealed AIST, two regimes are observed each following a rather Arrhenius-like temperature dependence; one at low heating rates $\vartheta$ up to 5,000 K/s and the other at heating rates above 10,000 K/s after a "kink" near 7,500 K/s. Upon the transition from the low to the high heating rate regime, the Kissinger activation energy drops by more than threefold from $3.06 \pm 0.05$ eV and $2.92 \pm 0.04$ eV to $0.89 \pm 0.16$ eV and $0.92 \pm 0.09$ eV for the as-deposited and pre-annealed phase, respectively. The kink constitutes a significant change in the crystallization process whose origin may not be deducted directly from the $T_p$ data presented in the Kissinger plot. A similar kink has been previously observed in the PCM $Ge_2Sb_2Te_5$ from cell voltage measurements [30] as well as from calorimetric measurements [31]. In addition, in the lower heating rate regime, the as-deposited phase shows a slight increase in activation energy $E_k$ with increasing $T_p$ that apparently is removed by pre-annealing. Pre-annealing on the other hand, generally shifts $T_p$ to higher temperatures indicative of a decreased crystallization speed. To understand the features observed in crystallization kinetics, i.e. the kink, the $T_p$-increase as well as the slight increase in $E_k$, we put forward a coherent view on crystallization kinetics, glass dynamics and the changes in the amorphous structure, as discussed below.

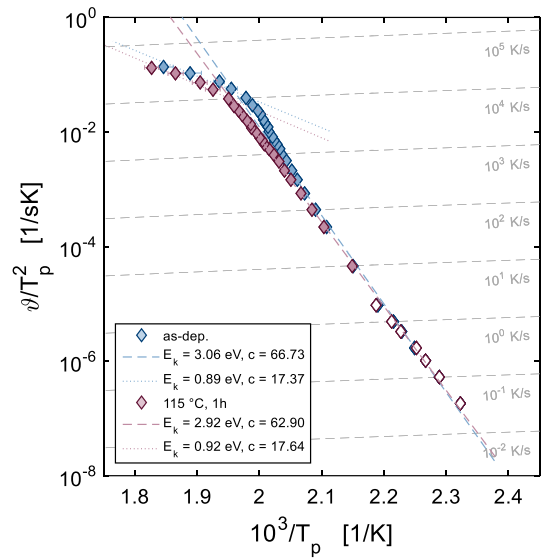

Figure 1: Kissinger plot of the peak temperature of crystallization $T_p$ of as-deposited and pre-annealed AIST obtained at constant heating rate $\vartheta$. From the lowest heating rate of $3.3 \times 10^{-2}$ K/s to about 5,000 K/s, $T_p$ follows a rather Arrhenius-like behavior indicative of a glassy phase. At higher heating rates, the activation energy (slope) shifts and decreases by more than threefold. Annealing causes the $T_p$ values to shift to higher temperatures, which indicates slower crystallization kinetics whereby the activation energy $E_k$ is decreased significantly. At heating rates above 10,000 K/s the activation energy $E_k$ of as-deposited and pre-annealed AIST is the same within the error.

Beginning with glass dynamics, insights can be gained by analyzing the excess specific heat capacity $C_p^{exc}$ as measured by (F)DSC. Measurements of $C_p^{exc}$ for as-deposited AIST at various heating rates $\vartheta$ are given in **Figure 2**. For lower heating rates (below 50 K/s), the first calorimetric feature is the exothermic heat release that sets in at about 100 °C. Such an exothermic heat release is readily observed in hyperquenched glasses, i.e. when the initial cooling rate is much larger than the (re-)heating rate [32-34]. The exothermic heat release reaches a maximum (minimum in Figure 2 as exothermic is down), which would normally be followed by an endothermic signal of the actual glass transition. However, in AIST a direct observation of the calorimetric glass transition is not possible since it is obscured by crystallization (see Figure 2). When the heating rate increases above 50 K/s, an endotherm known as the shadow glass transition develops prior to the exothermic heat release [32-38]. The shadow glass transition is usually induced by pre-annealing hyperquenched (fragile) glasses [36]. As discussed below, the shadow glass transition in AIST can indeed also be induced by pre-annealing. Interestingly, however, Figure 2 shows that the shadow glass transition is induced by an increase in the heating rate $\vartheta$ instead of being introduced by pre-annealing. This behavior was previously observed in another PCM $Ge_2Sb_2Te_5$ [31] and in metallic glasses [39]. As the heating rate exceeds 100 K/s, a sign of the actual glass transition after the exothermic heat release becomes visible as the excess heat capacity becomes slightly positive, i.e. endothermic. However, the full glass transition is still obscured by crystallization. Upon increasing the heating rate further, the shadow glass transition increases in size and dominance and the exothermic heat release becomes less pronounced while the actual glass transition becomes less obscured by crystallization.

The exothermic heat release should decrease and ultimately vanish when the (re-)heating rate is equal to the initial cooling rate [40, 41] as it is observed at 5,000 K/s. The vanishing of the exothermic heat release indicates that this heating rate is close to an effective cooling rate that can be attributed to the sputtering process [42]. This leads to the merging of shadow glass transition and the actual glass transition that now constitutes a clear single endothermic peak at heating rates of 5,000 K/s and above, where the endothermic signal is most pronounced. Note that at this heating rate, the $T_p$ values in the Kissinger plot start to deviate largely from the rather Arrhenius-like temperature dependence observed at low heating rates.

The coincidence of the change in calorimetric (F)DSC features with the kink in the Kissinger plot indicates that the glass transition is interwoven with this pronounced change in crystallization kinetics. Since the crystallization of AIST has been previously proposed to occur either from the glassy phase [17] or the UCL [43, 44], further analysis of the crystallization behavior is needed to explain the effect of pre-annealing on the crystallization kinetics.

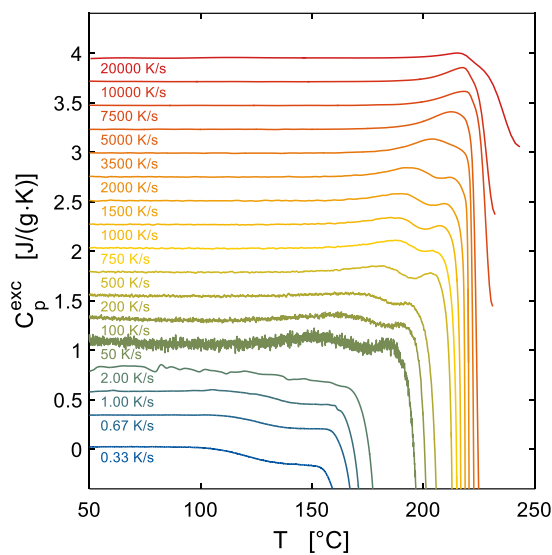

Figure 2: Combined (F)DSC excess specific heat capacity $C_p^{exc}$ curves for as-deposited AIST at various heating rates $\vartheta$. At low heating rates, only the exothermic heat release due to structural relaxation prior to crystallization is visible. At heating rates above 50 K/s a shadow glass transition endotherm develops prior to the exothermic heat release. Also at about 100 K/s and higher, the upward tilt of the actual glass transition starts to become endothermic just before crystallization. When the heating rate is increased, the shadow and the actual glass transition start to merge until they appear as one process at heating rates of 5,000 K/s and higher.

The effect of pre-annealing at different temperatures for one hour on $C_p^{exc}$ of AIST is presented in **Figure 3**. The DSC data of the as-deposited glass shows the aforementioned enthalpy relaxation exotherm with a minimum near 145 °C followed by crystallization at around 155 °C. Upon pre-annealing at a temperature of 55 °C or higher, some of the enthalpy is released during annealing and the exotherm progressively vanishes as shown in Figure 3. When the annealing temperature becomes 85 °C or higher, the endotherm of the shadow glass transition develops prior to the exotherm. Furthermore, the onset of crystallization progressively increases with the annealing temperature. This is in line with the increase in $T_p$ already observed in Figure 1, indicative of a reduced crystallization propensity.

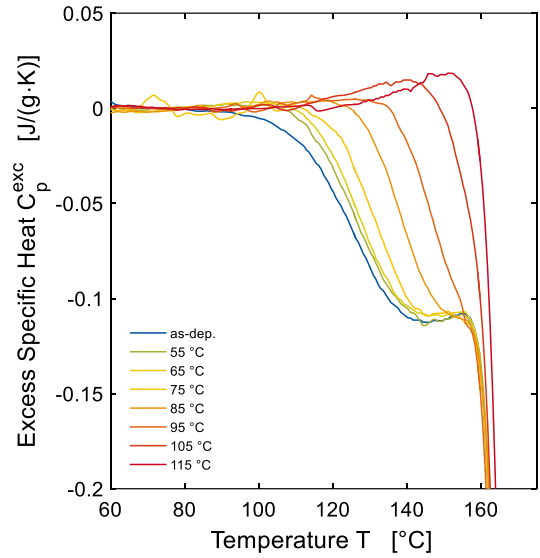

Figure 3: The effect of isothermal annealing for one hour on as-deposited AIST as reflected by the excess specific heat capacity $C_p^{exc}$ (endothermic is up). In the as-deposited phase, only the exothermic heat release caused by structural relaxation of the glassy phase is visible. At around 145 °C the enthalpy relaxation reaches its maximum (minimum in the Figure) before crystallization is initiated at about 155 °C. By annealing up to 95 °C, the relaxation exotherm is reduced continuously while the onset temperature of crystallization is not changed notably. When annealed at higher temperatures, the enthalpy relaxation endotherm is removed completely and is replaced by a shadow glass transition endotherm. Also the onset of crystallization is shifted to higher temperatures indicative of a decreased crystallization speed upon annealing.

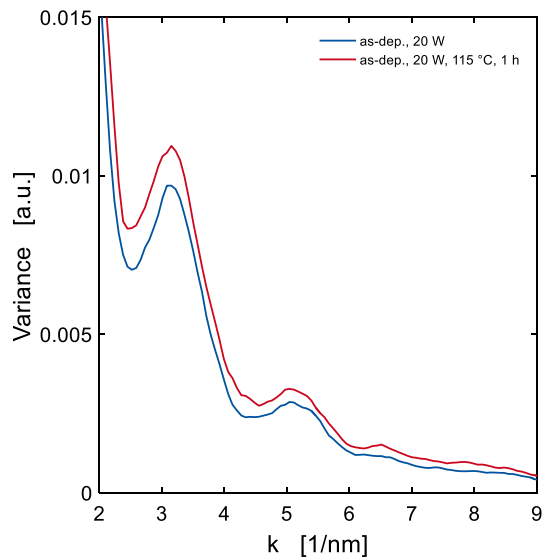

Figure 4: Fluctuation electron microscopy (FEM) variance for as-deposited and pre-annealed AIST. Pre-annealing at 115 °C for 1 h increases the variance, which is a clear sign of increased medium-range order (MRO). An increased FEM variance (MRO) of the PCM AIST has been linked to an increased tendency of nucleation before [45, 46].

Importantly, Figure 3 indicates that crystallization occurs while the system is still undergoing the exothermic enthalpy release. This implies that the system is still relaxing and therefore must still be out of thermodynamic equilibrium i.e. still be a glass when crystallization occurs. This suggests that the system crystallizes from the glassy state when low heating rates are applied. This is also consistent with the observation that the crystallization kinetics is slowed down upon annealing (Figure 1 and 3). Indeed, annealing would increase the glass viscosity and lower both the nucleation rate and growth velocity, thereby leading to an increase in $T_p$ as is observed experimentally in Figure 1. If crystallization instead occurred from the undercooled liquid phase (UCL), pre-annealing would not affect the viscosity $\eta$ as the UCL would behave in an ergodic manner and thus the crystal growth velocity $v_g$ would remain unchanged. However, the crystallization peak temperature could be influenced by a change in the subcritical nuclei distribution [47]. A reduced subcritical nuclei distribution could lead to a less pronounced nucleation tendency, so-called fading (the opposite of so-called priming for an increase in subcritical nuclei distribution) [48]. Hence, an estimate of the subcritical nuclei distribution should be performed to determine whether annealing promotes or hinders the nucleation tendency.

For amorphous AIST, the nucleation tendency has been shown to correlate with amount of the medium range order (MRO) as obtained from the variance in fluctuation electron microscopy (FEM) measurements [45, 46]. So, if AIST was crystallizing from the UCL, pre-annealing could lead to higher $T_p$ values as observed in Figure 1 and 3, if pre-annealing decreased the FEM variance. Measurements on the FEM variance, and thus the amount of MRO, of as-deposited and pre-annealed AIST are presented in **Figure 4**. Here an increase in FEM variance is obtained, which therefore should correlate with an increased nucleation tendency. An increased nucleation tendency should thus lead to lowered $T_p$ values. By contrast, the $T_p$ values are increased by pre-annealing which is inconsistent with crystallization from the UCL. These results therefore confirm that crystallization occurred from the glassy phase, since $T_p$ values decrease despite the increase in nucleation tendency suggested by FEM. Indeed, it is well-known that pre-annealing at a temperature below the glass transition temperature $T_g$ (or to be precise below its fictive temperature $T_f$ [21]) induces aging by structural relaxation whereby the viscosity $\eta$ increases [49] which in turn decreases the crystal growth velocity $v_g$. The increase in nucleation tendency is compensated by this decrease in crystal growth velocity $v_g$.

Our results show that the glass transition temperature $T_g$ of AIST must be larger than the pre-annealing temperature of 115 °C, which raises the question what the actual glass transition temperature is. In an attempt to estimate the $T_g$ of AIST, the exothermic heat release visible in $C_p^{exc}$ of the as-deposited phase is compared to that of hyperquenched well known glass formers as proposed in Ref. [50] and shown in **Figure 5** (exothermic is up). Here, the maximum enthalpy relaxation occurs at a temperature ranging from 0.88 to 0.96 $T_g$. An estimate of the glass transition temperature of AIST is then obtained by choosing the value of $T_g$ so that it is aligned with the average of these extrema at 0.92. This yields an estimate for the $T_g$ of AIST of 182.5 °C with an estimated uncertainty of ± 20 °C. As also shown in Figure 5, a glass transition temperature of 115 °C or below would be clearly too low, as the materials would still release enthalpy that was stored in the glassy phase at temperatures above the glass transition temperature and well in the UCL region, where ergodicity should easily be reached. Utilizing this estimated glass transition temperature, a value for the fragility $m$ of 103 is obtained by fitting the Mauro-Yue-Ellison-Gupta-Allan (MYEGA) model [51] to the high-temperature viscosity data of AIST from Ref. [44] (see Supplementary Information (SI)), which can describe the data satisfyingly well assuming a single fragility. As a consistency test of the estimated $T_g$ value, the expected glass transition temperature of 202 °C at a cooling rate $\vartheta_c$ of 5,000 K/s (fictive temperature) is calculated by the Moynihan method [42, 52, 53],

$$T_f(\vartheta_c) = T_g \cdot \left(1 - \frac{1}{m} \cdot \log_{10}\left(\frac{\vartheta_c}{\vartheta_c^s}\right)\right)^{-1} \qquad (2)$$

where $\vartheta_c^s$ is the standard cooling rate of 20 K/min. As noted above at an heating rate of 5,000 K/s the exothermic heat release has ceased which indicates that this heating rate is similar to an effective cooling rate of the sputtering process [42]. So the glass transition temperatures of cooling and (re-)heating should be equal [40, 54]. This value is in line with the onset temperature near 190 °C observed in Figure 2, although this onset could be underestimated by the fact that the rise of the endotherm of the glass transition is cut short by crystallization and thus should show a higher temperature comparable to the value calculated here.

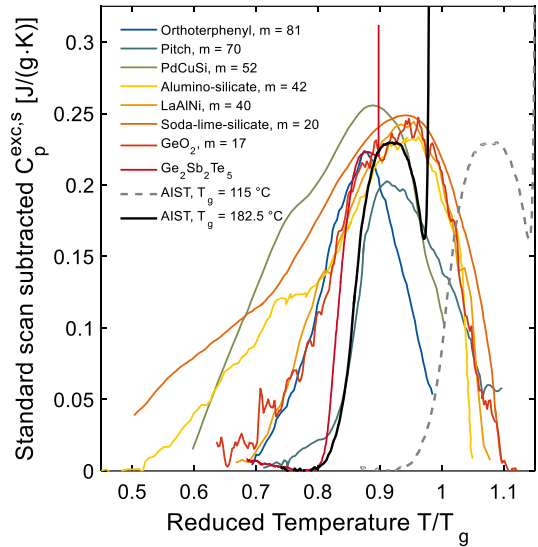

Figure 5: Standard scan subtracted excess specific heating rate $C_p^{exc,s}$ of several hyperquenched glasses as a function of temperature scaled by the glass transition temperature $T_g$ for orthoterphenyl [50], pitch [55], $Pd_{77.5}Cu_6Si_{16.5}$ [56], $La_{55}Al_{25}Ni_{20}$ [34], basaltic fiber [57], soda lime-silicate [58], $GeO_2$ [59] and (crystalline rescan subtracted) $Ge_2Sb_2Te_5$ [31]. The exothermic heat peak is found in the range from 0.88 to 0.96 of $T_g$. Now, the $T_g$ for AIST is chosen so that its exothermic heat release peak in $C_p^{exc}$ turns out at 0.92 of $T_g$, which yields to a value of 182.5 °C.

## 3. Conclusion

The combined results on crystallization kinetics (Figure 1) and glass dynamics (Figures 2 and 3) together with the measurements on the MRO (Figure 4) lead to the conclusion that at heating rates up to 5,000 K/s the crystallization of the PCM AIST is incompatible with crystallization from the UCL, but instead it crystallizes directly from a glassy phase. This conclusion is confirmed by the Arrhenius-like temperature dependence (Figure 1) and the fact that the crystallization speed decreases (increase in $T_p$, see also Figure 1) upon pre-annealing. The glass transition temperature $T_g$ of AIST was estimated to be about 182.5 °C. The estimated value for $T_g$ means that at conventional heating rates, crystallization initiates about 27.5 °C below the glass transition, which highlights the large tendency of the PCM AIST to crystallize where the atomic mobility should hinder significant crystallization. Note that recent studies observed that PCMs undergo pronounced $\beta$-relaxations prior to crystallization [60], where activated fast local structural rearrangement may play a non-trivial role in facilitating crystallization. In addition, this study will help understand the discrepancy in reported $T_g$ values and crystallization behaviors reported in the literature [17, 43, 44, 61].

## 4. Experimental Section/Methods

Powders for conventional differential scanning calorimetry (DSC) and ultrafast or flash DSC (FDSC) measurements of the phase change material (PCM) $Ag_4In_3Sb_{67}Te_{26}$ are produced from a stoichiometric target by magnetron sputter deposition at a base pressure of $3\times10^{-6}$ mbar. The composition was confirmed via Scanning Electron Microscopy (SEM) in a FEI Helios Dual Beam FIB. The excess specific heat capacity $C_p^{exc}$ was obtained in a PerkinElmer Diamond DSC and a Mettler-Toledo Flash DSC 1 by subtracting the crystalline rescan from the initial scan. The as-deposited and pre-annealed AIST samples were measured on the same sensor in FDSC. The melting onset of pure Indium was used to calibrate the temperature at a constant heating $\vartheta$. Fluctuation electron microscopy (FEM) was obtained at an FEI Titan G2 80-300 STEM on 30 nm thick AIST samples sputtered directly onto amorphous carbon films supported by copper grids. For both phases at least 1,000 diffraction patterns were acquired from which the FEM variance was calculated. For the calibration of the detector for real space and reciprocal space readings a silicon reference was used. The electron beam diameter (probe size) and an divergence angle of 0.35 mrad was obtained to be 1.36 nm (FWHM).

**Acknowledgements**
The authors acknowledge funding from the Deutsche Forschungsgemeinschaft (DFG) via the collaborative research center Nanoswitches (SFB 917). P.L. acknowledges funding from NSF-DMR grant No. 1832817. S.W. acknowledges the support of the DFG grant No. AOBJ670132.

## Author contributions
Julian Pries: Conceptualization, project administration, sample preparation, DSC and FEM investigation and data curation, formal analysis, writing-original draft, reviewing and editing. Julia Sehringer: FDSC investigation and data curation. Shuai Wei: Writing-reviewing and editing. Pierre Lucas: Writing-reviewing and editing. Matthias Wuttig: Supervision, writing-reviewing and editing. All authors have given approval to the final version of the manuscript.

## Financial and conflict of interest
The authors declare not conflict of interest.

J. Pries, J. Sehringer, S. Wei, P. Lucas, M. Wuttig*

**Glass Transition of the Phase Change Material AIST and its Impact on Crystallization**

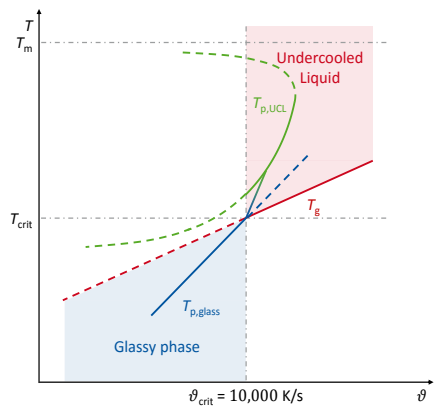

Calorimetric data on the excess heat capacity $C_p^{exc}$ and the crystallization peak temperature $T_p$ is combined to identify the amorphous state the phase change material AIST crystallizes from, i.e. the glassy phase or the undercooled liquid. When glass transition and crystallization interfere, the Kissinger activation energy $E_k$ drops by more than threefold, enabling rapid crystallization. The glass transition temperature $T_g$ is estimated to be about 182.5 °C. Hence, at conventional heating rates, AIST unconventionally crystallizes ∼ 27.5 °C below the actual glass transition.

Supporting Information

# Glass Transition of the Phase Change Material AIST and its Impact on Crystallization

*Julian Pries, Julia Sehringer, Shuai Wei, Pierre Lucas, Matthias Wuttig\**

**1. Fragility of AIST**

Literature data on the viscosity $\eta$ of (Ag,In)-doped $Sb_2Te$ (AIST) are only available above the melting point and reported in Ref. [1] for the liquid phase from 829 K to 1254 K. The melting temperature was also reported to be 807 K elsewhere [2]. Together with the estimated glass transition temperature $T_g$ from the main text of 182.5 °C, the single-fragility Mauro-Yue-Ellison-Gupta-Allan (MYEGA) model is fitted to the viscosity data. The MYEGA function is [3]

$$\log_{10} \eta = \log_{10} \eta_\infty + (12 - \log_{10} \eta_\infty) \frac{T_g}{T} \exp\left[\left(\frac{m}{12 - \log_{10} \eta_\infty} - 1\right)\left(\frac{T_g}{T} - 1\right)\right] \qquad (1)$$

where $m$ is the fragility, and $\eta_\infty$ is the limit of viscosity at infinite temperature. The fragility $m$ is defined by [4]

$$m = \left.\frac{\mathrm{d}\log_{10}\eta(T)}{\mathrm{d}\left(\frac{T_g}{T}\right)}\right|_{T=T_g}. \qquad (2)$$

For fitting the MYEGA model to the viscosity data, the glass transition temperature $T_g$ estimated in the main text was taken while $\eta_\infty$ and $m$ were chosen as free parameters. The fitting resulted in values for fragility $m$ and $\eta_\infty$ to be $103.3 \pm 0.3$ and $(6.73 \times 10^{-4} \pm 0.05 \times 10^{-4})$ Pas, respectively. The errors are only due to fitting. Note that since the viscosity data are only available at high temperature, the single-fragility fit corresponds to the high-temperature fragile liquid state. A fragile-to-strong transition is supposed to occur at about 20 % below the melting point, according to a recent structural study using femtosecond diffractions [5]; however, the lack of viscosity data in that temperature range does not allow for a proper fit for the strong liquid (low temperature) state.

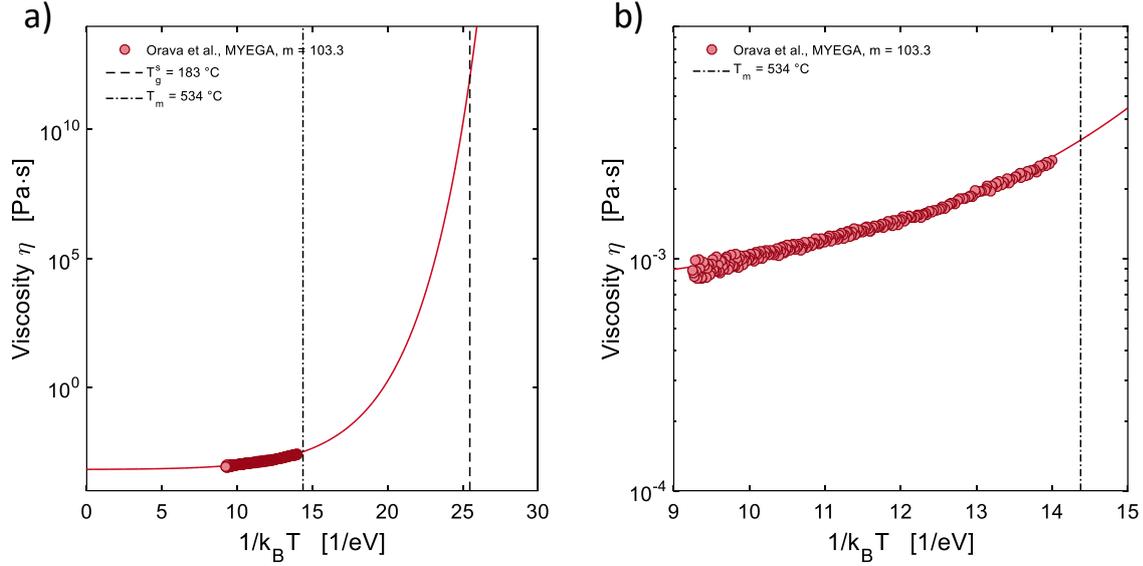

Figure 1: Viscosity as a function of inversely $k_B$-scaled temperature. Viscosity data taken from Ref. [1]. The melting temperature $T_m$ was reported in Ref. [2] and the glass transition temperature $T_g$ of 182.5 °C was estimated in the main text- Red line constitutes the MYEGA fit to the viscosity data, taking $T_g$ to be fixed and $\eta_\infty$ as a free parameter. b) is a close-up of a).